\documentstyle[twocolumn,aps,prl,epsfig]{revtex}
\begin{document}
\noindent Replacement cond-mat/0002383\\
{\bf Asymptotic Energy Decay in Inelastic Fluids}\\
J.A.G.~Orza, R.~Brito and M.H. Ernst\\
Dpto.~de F\'\i sica  Aplicada I, Universidad Complutense\\
 28040 Madrid, Spain. \\

The goal of the present publication is the  comparison 
of the two existing theoretical predictions \cite{BE,Ben-Naim}
for the long time  behavior  of the total energy $E(t)$ 
in a {\em large}  system of freely evolving 
inelastic hard spheres (IHS) with computer simulations of 
$N=50\,000$ hard disks. 

 (i) The first theory,  a mode coupling theory by Brito and Ernst \cite{BE}, 
predicts that $E \sim A\tau^{-d/2}$ for $d \geq 2$, where  $\tau (t) $ 
is the average number of collisions suffered by a particle in time $t$ 
($\tau$ is a nonlinear function of $t$, whose analytic form is unknown 
at large $t$) and  $A$ is a known coefficient, which  {\it depends} on the 
coefficient of restitution, $r$, and on the density, $\phi$. 

\begin{figure}
\epsfig{file=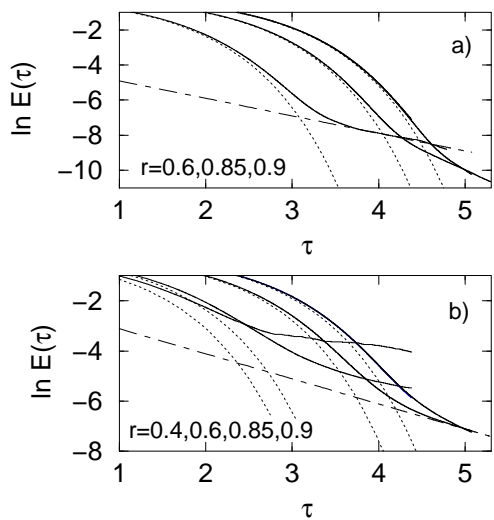,width=6.5cm}
\caption{Energy vs $\tau$  at
packing fractions $\phi=0.05$ (upper) and $0.4$ (lower)
at different inelasticities. The dotted lines
represent the short time behavior
(Haff's law), and the dashed-dotted lines the theory of [1].}
\end{figure}

Computer simulations of $\ln E(\tau)/E(0)$ are shown in Fig.~1,
where we have chosen units such that $E(0)=1$.
These plots confirm that the small fluctuation theory of
\cite{BE} gives at fixed $\phi$ quantitative predictions 
inside an $r-$window of width $\Delta r \simeq 0.1$, centered
around $r_0 \simeq  0.6,\, 0.75,\,0.8,\, 0.85$
at densities $\phi \simeq 0.05,\, 0.11,\,0.245,\,0.4$
respectively. At larger/smaller $r-$values, $E(\tau)$ decays 
faster/slower than the prediction of \cite{BE}. The location $r_0$ is 
determined by two balancing effects: sufficiently high density 
to suppress large relative density fluctuations which increase 
the mean overall collision frequency $d \tau /dt$ (compression 
of $\tau-$axis),  {\em competing} with sufficiently high 
inelasticity, which favors local inelastic collapses (a finite 
number of collisions in an infinitesimal time). Such collisions
merely increase the $\tau -$value (stretching of $\tau -$axis)
at fixed $t$ without advancing the $N$ particle dynamics.
At small density the relative density fluctuations  
become very large, as the clusters keep growing and  compactifying,
which invalidates the linear theory of \cite{BE}. 

(ii) Ben-Naim {\em et al.}  \cite{Ben-Naim}
show that the behavior of a fluid 
of inelastic hard rods is described by the 
totally inelastic (sticky) gas. Consequently, the energy decays
at long times as $t^{-2/3}$ in 1-D, and is independent of the 
coefficient of restitution. Moreover,  they conjecture that the total
energy $E$ for $ 2\leq d\leq 4$ decays as $ B t^{-d/2}$ with an
unknown coefficient $B$, {\em independent} of inelasticity.
The results of \cite{Ben-Naim} are only valid at 
asymptotically large times.

Our goal is to test this conjecture against simulations of
inelastic hard disks.  The energy decay when  plotted as 
$\log E$ versus $\log t$ (see Fig.~2), gives the misleading 
impression that our simulations have reached their 
asymptotic time dependence, and suggests that $ E \sim  C t^{-a}$ 
decays algebraically with a density-dependent exponent $a$, 
but with $a$ and $C$ {\it independent } of the dissipation, 
possibly corresponding to a sticky hard sphere fluid.  
At small densities (Fig.~2a), $a\simeq 1$, which offers 
partial support for the conjecture, while at higher 
densities (Fig.~2b) the analysis seems to show a smaller 
exponent. 

\begin{figure}
\epsfig{file=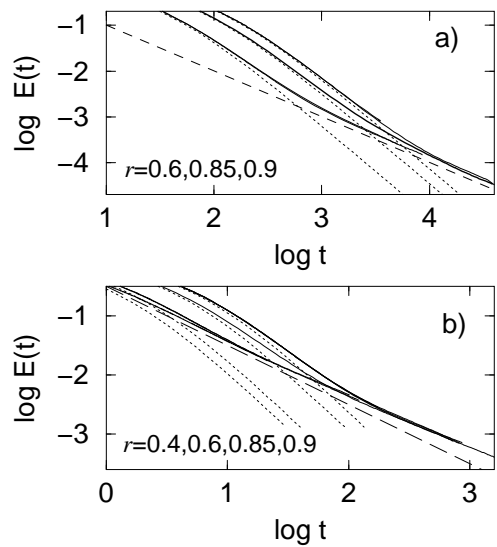,width=6.5cm}
\caption{Energy vs $t$ at 
packing fractions $\phi=0.05$ (upper) and $0.4$ (lower)
at different restitution coefficient $r$. The dotted lines 
represent the short time behavior 
(Haff's law). Long-dashed line with slope $-1$ is plotted 
for reference.}
\end{figure}

However, a more sensitive test is to
plot  $E(0)/E(t)$ versus $t$, and test whether the curves for different 
values of the restitution coefficient  $r$ become linear in $t$,
and tend to coincide for large times, i.e. become independent of $r$.
This is done in Fig.~3 for 
three different packing fractions $\phi=0.05,0.245,0.4$, each for a range
of $r$--values over a long time interval far beyond where Haff's 
law is valid. The results at low density can hardly be considered as
evidence for the conjecture. The  curves at 
intermediate density, $\phi=0.245$ show behavior 
that might be linear in $t$, but depends strongly on the degree of
inelasticity. Simulations at density $\phi=0.11$ show behavior similar 
to the ones for $\phi=0.245$.
The behavior at the highest density, at $\phi=0.4$, in the time
interval considered, looks roughly independent of the degree of 
inelasticity, but the curves show a tendency to diverge at later times.

\begin{figure}
\epsfig{file=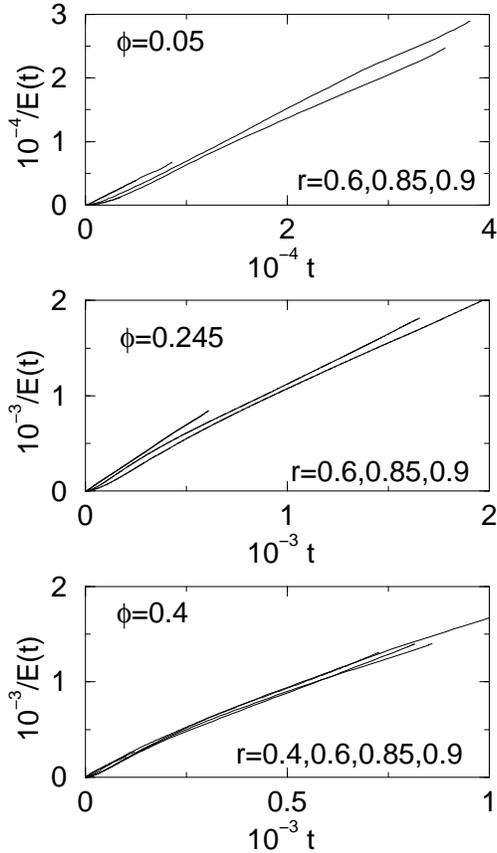,width=6.5cm}
\caption{Long time behavior of the inverse of the energy vs 
$t$  at packing fractions $\phi=0.05$ (upper), $\phi=0.245$
(middle)  and $\phi=0.4$ (lower) at different dissipations $r$. 
Short time behavior, $1/E(t)\sim(1+t/t_e)^2$, as  
described by Haff's law is still visible.}
\end{figure}

In general, the curves in Fig.~3 do not seem to have 
reached their asymptotic form, and are not conclusive 
enough to support or refute conjectures about the asymptotic 
$t$-dependence being independent of the degree of
inelasticity. 

Finally we observe that at asymptotically large times
there is an important distinction between {\em  small}
and thermodynamically {\em large} systems, for the following reason. 
The growth of patterns, i.e.~vortices and density clusters, is 
controlled by diffusive modes \cite{BE}, and typical diameters,
$L_v(\tau)$ of these patterns grow as $\sqrt\tau$. As soon
as the system size $L\sim  L_v(\tau)$,
patterns start to interfere with their periodic images, and there
occurs crossover to a {\em steady state}, which is fully 
determined by the (unphysical) periodic bounday conditions.
This is the case for asymptotically large times in small 
systems. In thermodynamically large systems this crossover 
is never reached. 

In {\em small} systems  we have observed that the energy 
decays as, $E(t) \sim C(N, r) /t^2$ for $t \to \infty$, 
with a coefficient  $C$ that depends on $N$ and $r$, 
and that differs in general from Haff's law. 
The law $E(t) \sim t^{-2}$ seems to hold for small 
systems in any dimension larger than 1 
(see also \cite{Chen,Trizac}).


\vspace*{-0.5cm} 
\begin{thebibliography}{9} 
\vspace*{-1.0cm} 
\bibitem{BE} R.~Brito and M.H.~Ernst, Europhys.~Lett. {\bf 43}, 497 (1998).
\bibitem{Ben-Naim} E.~Ben-Naim, S.Y.~Chen, G.D.~Doolen and S.~Redner, 
Phys.~Rev.~Lett. {\bf 83}, 4069 (1999).
\bibitem{Chen}  S.~Chen, Y.~Deng, X.~Nie and Y.~Tu,
Phys.~Lett.~A, {\bf 269}, 218 (2000). 
\bibitem{Trizac} E.~Trizac and  A.~Barrat, cond-mat/0006106. 
\end{thebibliography}
\end{document}